\documentclass{elsart}

\usepackage{amsmath}                   
\usepackage{amssymb}                   
\usepackage{graphicx}                  
\usepackage{subfigure}
\usepackage{rotate}
\begin{document}


\begin{frontmatter}
\journal{Nucl. Instr. Meth. A}

\title{
The Lidar System of the\\
Pierre Auger Observatory
}

\author[Columbia]{S.Y.~BenZvi},
\author[Torino]{R.~Cester},
\author[Torino]{M.~Chiosso},
\author[Columbia]{B.M.~Connolly},
\author[NGP]{A.~Filip\v{c}i\v{c}},
\author[UTN]{B.~Garc\'{\i}a},
\author[GranSasso]{A.~Grillo},
\author[Napoli]{F.~Guarino},
\author[NGP]{M.~Horvat},
\author[Aquila1]{M.~Iarlori},
\author[Aquila2]{C.~Macolino},
\author[UNM]{J.A.J.~Matthews},
\author[Tandar]{D.~Melo},
\author[Torino]{R.~Mussa},
\author[Utah]{M.~Mostaf\'a},
\author[CEILAP]{J.~Pallota},
\author[Aquila2]{S.~Petrera},
\author[Columbia]{M.~Prouza},
\author[Aquila1]{V.~Rizi},
\author[PSU]{M.~Roberts},
\author[UTN,Roma]{J.R.~Rodriguez Rojo},
\author[Aquila2]{F.~Salamida},
\author[Auger]{M.~Santander},
\author[Torino]{G.~Sequeiros},
\author[Torino]{A.~Tonachini},
\author[Napoli]{L.~Valore},
\author[NGP]{D.~Veberi\v{c}},
\author[Columbia]{S.~Westerhoff},
\author[NGP]{D.~Zavrtanik}, and
\author[NGP]{M.~Zavrtanik}

\address[Auger]{Pierre Auger Southern Observatory, Av. San Martin Norte 304, 5613 Malarg\"ue,
                Prov. de Mendoza, Argentina}
\address[Tandar]{Laboratorio Tandar, Comisi\'on Nacional de Energ\'{\i}a At\'omica,
                 Av. del Libertador 8250, Buenos Aires, Argentina}
\address[UTN]{Universidad Tecnol\'ogica Nacional, Regionales Mendoza y San
             Rafael; CONICET, Rodr\'{\i}guez 273 Mendoza, Argentina}
\address[CEILAP]{CEILAP-CITEFA/CONICET, Villa Martelli, Prov. Buenos Aires, Argentina}
\address[Aquila1]{CETEMPS/Dipartimento di Fisica dell'Universit\`a de l'Aquila and INFN, Gruppo
                  Collegato, Via Vetoio, I-67010 Coppito, Aquila, Italy}
\address[Aquila2]{Dipartimento di Fisica dell'Universit\`a de l'Aquila and INFN, Gruppo
                  Collegato, Via Vetoio, I-67010 Coppito, Aquila, Italy}
\address[GranSasso]{Laboratori Nazionali del Gran Sasso, Strada Statale 17/bis Km 18+910,
                    I-67010 Assergi (L'Aquila), Italy}
\address[Napoli]{Dipartimento di Fisica dell'Universit\`{a} di Napoli and
                 Sezione INFN, Via Cintia 2, I-80123 Napoli, Italy}
\address[Roma]{Dipartimento di Fisica dell'Universit\`{a} di Roma II ``Tor Vergata'' and Sezione 
               INFN, Via della Ricerca Scientifica, I-00133 Roma, Italy}
\address[Torino]{Dipartimento di Fisica Sperimentale dell'Universit\`a di Torino
                 and Sezione INFN, Via Pietro Giuria, 1, I-10125 Torino, Italy}
\address[NGP]{University of Nova Gorica, Laboratory for Astroparticle Physics, Vipavska 13,
              POB 301, SI-5000 Nova Gorica, Slovenia}
\address[Columbia]{Columbia University, Department of Physics and Nevis Laboratories,
                   538 West $\it 120^{th}$ Street, New York, NY 10027, USA}
\address[PSU]{Pennsylvania State University, Department of Physics,
              104 Davey Lab, University Park, PA 16802, USA}
\address[UNM]{University of New Mexico, Department of Physics and Astronomy, Albuquerque, 
              NM 87131, USA}
\address[Utah]{University of Utah, 115 S. 1400 East \# 201, Salt Lake City, UT 84112, USA}


\begin{abstract}

The Pierre Auger Observatory in Malarg\"ue, Argentina, is designed to study the
origin of ultrahigh energy cosmic rays with energies above $10^{18}$\,eV.  The energy 
calibration of the detector is based on a system of four air fluorescence detectors. 
To obtain reliable calorimetric information from the fluorescence stations, the
atmospheric conditions at the experiment's site need to be monitored continuously
during operation.  One of the components of the observatory's atmospheric monitoring 
system is a set of four elastic backscatter lidar stations, one station at each of 
the fluorescence detector sites.  This paper describes the design, current
status, standard operation procedure, and performance of the lidar system of the 
Pierre Auger Observatory.

\end{abstract}

\begin{keyword} 
Ultrahigh energy cosmic rays; air fluorescence detectors; atmospheric monitoring; lidar
\PACS 07.60.-j \sep 42.68.Wt \sep 92.60.-e \sep 96.50.sd
\end{keyword}

\end{frontmatter}

\section{Introduction}\label{sec:intro}

The Pierre Auger Observatory in Malarg\"ue, Argentina, the world's largest facility 
to detect ultrahigh energy cosmic rays with energies above $10^{18}$\,eV, 
is currently nearing completion.  Data collection started in 
January 2004, and the observatory has already accumulated an
ultrahigh energy cosmic ray data set of unprecedented size.

Due to their extremely low flux, primary cosmic rays at ultrahigh energies 
cannot be observed directly.  Rather, they
enter the atmosphere and interact with air molecules, inducing a cascade 
of secondary particles, called an extensive air shower.  The properties of the 
primary particle (energy, arrival direction, and chemical composition) have to 
be deduced from indirect measurements of the extensive air shower. 

Two detector types have traditionally been used to measure air showers: surface 
detectors, which detect the particles of the air shower cascade that reach the ground;
and air fluorescence detectors, which make use of the fact that the particles in 
the air shower excite air molecules, causing UV fluorescence.  Observing this UV
light from air showers with photomultiplier cameras allows us to image
the air shower development and obtain a nearly calorimetric energy estimate of the
shower.  For both detector types, the atmosphere acts as the calorimeter and provides the 
large detector volume needed to measure the small cosmic ray flux at the highest energies.  

The Pierre Auger Observatory is a hybrid detector which combines a large 
surface detector (SD) array of area $3000\,\mathrm{km}^{2}$ and a system 
of four fluorescence telescopes (FD) at the same site.  While the SD provides 
the well-defined aperture and the $100\,\%$ duty cycle needed to achieve
high statistics, the FD provides, for a subset of showers, the calorimetric 
information needed to calibrate the SD. 

For the calibration to be meaningful, the properties of the calorimeter, 
{\it i.e.} the atmosphere, must be well-known.  Particulates in the 
atmosphere, Cherenkov radiation and the weather need to be factored into 
this calibration.  Fluorescence light from 
air showers is affected by the absorption and scattering on molecules and
aerosols, and, in addition, some fraction of the Cherenkov radiation from 
the relativistic particles in the air shower is scattered into the 
detector.  This Cherenkov light cannot be separated from the fluorescence 
light, and so the contamination by Cherenkov light needs to be modeled and 
subtracted to give an accurate energy estimate for the cosmic ray shower. 
Moreover, the light transmission properties of the atmosphere are not 
constant over the large detector volume of the Pierre Auger Observatory, 
and equally important, they are not constant in time.  Changing weather 
conditions inevitably mean changing transmission properties.  Consequently, 
the properties of the atmosphere need to be monitored continuously.

The Pierre Auger Observatory has an extensive program to monitor the 
atmosphere within the FD aperture and measure atmospheric attenuation 
and scattering properties in the 300 to 400\,nm sensitivity range of 
the FDs.  A summary of the atmospheric monitoring system can be found 
in~\cite{Cester:2005}.  Within this system, a central role is played 
by a system of four elastic\footnote{Here, the term {\it elastic} refers to the
light scattering process.  In elastic lidar applications, the return signal
is measured at the same wavelength as the original laser signal.}
backscatter lidar ({\it li}ght {\it d}etection 
{\it a}nd {\it r}anging) stations, one at each fluorescence site.
system is currently under construction, with three out of four stations
operating.  At each lidar station, a high-repetition UV laser sends 
short laser light pulses into the atmosphere in the direction of interest. 
The backscattered signal is detected as a function of time by photomultiplier 
tubes at the foci of parabolic mirrors.  Both the laser and the mirrors are 
mounted on a steering frame that allows the lidar to cover the full azimuth
and elevation of the sky.

During each hour of FD data taking, the four lidars perform a routine scan of 
the sky over each FD.  The data provide information about the height and 
coverage of clouds as well as their depth and opacity, and the local aerosol 
scattering and absorption properties of the atmosphere.  In addition to this 
routine operation, the lidar system is used for real time monitoring of the 
atmospheric homogeneity between the FDs and selected cosmic ray events.  
For example, if a high energy ``hybrid'' event is observed with the SD and 
one or more FDs, the routine scan is interrupted and, within 2 to 4 minutes 
of the event detection, the lidar scans the atmosphere in the vicinity of the 
air shower reported by the FD.  This procedure is called "shoot-the-shower" (StS),
and allows for a rejection of events where the light profile from the track 
is distorted by clouds or other aerosol non-uniformities that are not 
characterized well by the average hourly aerosol measurements.
Both light reflection and opacity can distort the light profile.

This paper describes the design, standard operation procedure, and performance 
of the lidar system of the Pierre Auger Observatory.  It is organized as follows.  
After a brief description of the SD and FD detectors (Section\,\ref{sec:auger}), 
we give a summary of the relevant atmospheric parameters (Section\,\ref{sec:atmo}). 
Section\,\ref{sec:hardware} describes the current lidar hardware.  In 
Section\,\ref{sec:ops}, the daily operating procedure is summarized, including a 
description of the routine scan and the shoot-the-shower operation.  First results 
of the analysis of atmospheric parameters like cloud height and extinction are 
summarized in Section\,\ref{sec:analysis}.  A section on inelastic Raman lidar 
measurements and a summary conclude the paper.

\begin{figure}[t]
  \centering
  \includegraphics[width=.7\textwidth]{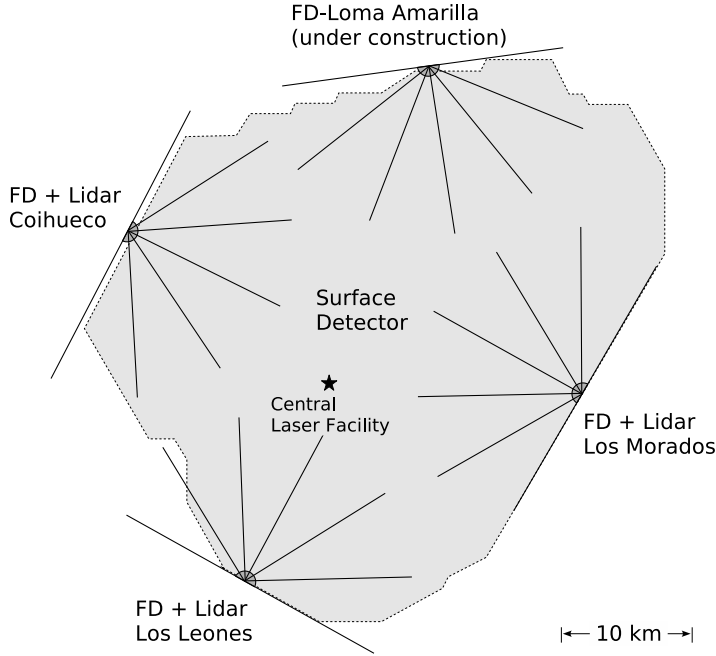}
  \caption{\it Schematic layout of the Pierre Auger Observatory.}
  \label{f:observatory}
  \vskip1cm
\end{figure}

\section{The Pierre Auger Observatory}\label{sec:auger}

In its final stage, the surface detector array of the Pierre Auger Observatory 
will comprise 1600 water Cherenkov detector tanks, deployed over an area of 
$3000\,\mathrm{km}^{2}$ on a triangular grid with 1500\,m grid spacing.  Each 
of these cylindrical plastic tanks is 1.5\,m tall with a $10\,\mathrm{m}^{2}$ base,
and filled with 12,000\,liters of purified water within a Tyvek liner.
Three $9^{\prime\prime}$ photomultiplier tubes in the tanks measure the Cherenkov light
produced by the particles of the extensive air shower cascade traversing the tank water.

The SD measures the particle density as a function of distance to the shower core, 
defined as the location where the primary particle would have hit the ground had it not 
showered in the atmosphere.  The SD array therefore measures the lateral profile of 
the shower at ground level.  The SD operates with a nearly $100\,\%$ duty cycle.

For a fraction of the showers, the SD measurement is complemented by a measurement
of the longitudinal profile of the shower in the atmosphere.  This information is provided 
by four fluorescence detectors.  Unlike Cherenkov light, which is strongly forward-peaked,
fluorescence light is emitted isotropically.  The fluorescence signal from the air shower 
is small compared to other light sources, but during dark, moonless nights, about 
$10\,\%$ of the total year, the fluorescence light can be distinguished from the night 
sky background. 

The FD comprises four detector stations overlooking the SD from the periphery 
(Fig.\,\ref{f:observatory}).  At present, the sites at Los Leones, Coihueco, and Los 
Morados are completed and fully operational, while the fourth site at Loma Amarilla 
is under construction.  Each site contains six bays, and each bay encloses a UV telescope 
comprised of a spherical light-collecting mirror, a photomultiplier camera in the focal 
surface, and a UV transmitting filter in  the aperture.  The mirrors have a radius 
of curvature of 3.4\,m and an area of about $3.5\times 3.5\,\mathrm{m}^{2}$.  The camera 
consists of 440 photomultipliers with a hexagonal bialkaline photocathode, arranged in a 
$20\times 22$ array.  Each camera has a field of view of $30.0^{\circ}$ in azimuth and
$28.6^{\circ}$ in elevation, covering an elevation angle range from $1.6^{\circ}$ to
$30.2^{\circ}$ above horizon.  
To reduce optical aberrations, the FD telescopes use Schmidt optics with a circular 
diaphragm of diameter 2.2\,m placed at the center of curvature of the mirror.  Coma 
aberrations are minimized by use of a refractive corrector ring at the telescope aperture.

The FD measures the amount of fluorescence light as a function of slant depth in the 
atmosphere.  This is then converted to the longitudinal profile of the shower, {\it i.e.} 
the number of charged particles as a function of slant depth.  The integral under the 
longitudinal profile is related to the electromagnetic energy of the shower, which can
be converted to the total energy of the primary particle.  This method of energy 
determination is nearly calorimetric and mostly independent of the assumed particle
interaction model.

The observatory started operation in hybrid mode in January 2004.  First physics results,
among them a first energy spectrum, were published based on data taken between 
January 2004 and May 2005 (see~\cite{Mantsch:2005} for a summary and further references).
The strength of the observatory is its ability to 
calibrate the SD using FD-SD hybrid events to find an empirical relationship between 
the energy determined by the FD and the particle density on the ground measured by the SD.  
Detailed knowledge of the atmosphere is a crucial requirement to make this energy calibration
work, as the accuracy of the energy calibration depends on an accurate measurement of the
molecular atmosphere and the atmosphere's aerosol content.

\section{Atmospheric Parameters}\label{sec:atmo}

The FD observes fluorescence light produced along the trajectory of the air shower
in the atmosphere.  The fluorescence signal is roughly proportional to the number of
particles in the shower cascade and therefore to the energy of the primary cosmic ray particle.
The fluorescence light traverses a path $l$ from the point of origin to the detector,
and the detected light intensity $P_{d}$ is weakened due to scattering and absorption 
on molecules or aerosols in the atmosphere.  It is related to the light intensity
at the place of its production, $P_{0}$, by

\begin{equation}
P_{d} = P_{0} \exp\left[-\int_{0}^{l} \alpha(r)r\right] \mathrm{d}r
      = P_{0} \exp\left[-\tau(l)\right] \hspace{10pt},
\end{equation}

where $\alpha(r)$ is the volume extinction coefficient, and $\tau(l)$ is the
optical depth to the shower point at distance $l$.  To obtain $\tau(l)$, 
$\alpha$ is integrated along the fluorescence light path between the detector and 
a given point on the shower.

For atmospheric fluorescence measurements, we need to account for two scattering 
processes:  Rayleigh scattering from atmospheric molecules and scattering from
low-altitude aerosols.  
Both molecules and aerosols in the atmosphere predominantly scatter, rather
than absorb, UV photons.  Some absorption does occur due to the presence of
ozone in the atmosphere and because the single scatter albedo of the aerosols 
is typically slightly less than unity, but these effects are small.  In general, 
fluorescence photons that are scattered by the atmosphere do not contribute to 
the reconstructed fluorescence light signal, although a small multiple scattering 
correction can be made for those that do~\cite{Roberts:2005}.  In this paper we 
will use the term ``attenuation'' when photons are scattered in such a way that 
they do not contribute to the light signal recorded at an FD or lidar.

The attenuation can then be described in terms of optical depths $\tau_{mol}$ for 
the molecular component and $\tau_{aer}$ for the aerosols.  Neglecting multiple 
scattering, the transmission of light through a vertical column of atmosphere 
can be expressed as

\begin{equation}
P_{d} = P_{0} \exp\left[-\tau_{mol}(l)-\tau_{aer}(l)\right] \hspace{10pt}.
\end{equation}

The parameters $\tau_{mol}$ and $\tau_{aer}$ need to be measured to relate the 
measured light intensity $P_{d}$ to $P_{0}$ and thus the shower energy.

While absorption and scattering processes of the molecular atmosphere are well 
understood~\cite{Bucholtz:1995}, the influence of the aerosol component is not.  
In principle, if one assumes that the aerosols are spherical with a known or 
assumed size distribution, then the light scattering can be described analytically 
by Mie scattering theory~\cite{Mie:1908,Hulst:1957,McCartney:1976}.  In practice, 
however, aerosols vary in size and shape, and the aerosol content changes on short 
time scales as wind lifts up dust, weather fronts pass through, or rain removes 
dust from the atmosphere.  

The aerosol optical depth $\tau_{aer}$ therefore needs to be monitored 
on at least an hourly basis during each night of FD observation over the entire
detection volume, which for the Pierre Auger Observatory corresponds to a ground
area of $3000\,\mathrm{km}^{2}$ up to a height of about 15\,km, the limit of the
detector aperture.  

At each FD, the local lidars provide the aerosol scattering coefficient, 
$\alpha_{aer}$, as a function of height $h$.  The value of $\alpha_{aer}(h)$ determines 
the amount of Cherenkov light scattered from the extensive air showers.  The 
integral of $\alpha_{aer}$ from the FD height to $h$ gives the vertical aerosol 
optical depth, $\tau_{aer}(h)$, which determines the transmission loss of 
light from each segment of the cosmic ray track to the FD.  A method to obtain 
$\tau_{aer}$ from lidar scans is described in detail in~\cite{Veberic:2002}.  In 
addition, the lidars provide the height and location of clouds and the optical 
depth inside the lowest cloud.  This analysis is described in 
Section\,\ref{sec:analysis}.

There is some overlap between the primary tasks of the lidars and a separate 
device for atmospheric monitoring at the Pierre Auger Observatory, the so-called 
Central Laser Facility (CLF). The CLF is situated in the center of the SD array.
From the CLF, a pulsed UV laser beam is directed into the sky, providing a test 
beam which can be observed by the FD.  The design and performance of the CLF is 
described in more detail in~\cite{Fick:2006}.

Choosing a one-hour time bin for the characterization of the
atmosphere is to some extent an arbitrary decision and a compromise, given that
the operation of lasers at the site interferes with the FD measurement and causes 
detector deadtime (see Section~\ref{subsec:opstypical}).  Measurements of atmospheric
parameters over the first years of data taking indicate that parameters do not
change on much shorter time scales.  However, this assumption needs to be tested 
with more data.  Crucial for this test will be data from the shoot-the-shower 
mode of lidar operation, taken within minutes of selected cosmic ray events, which 
can be compared to the hourly measurement.

\begin{figure}[t]
  \centering
  \includegraphics[width=.75\textwidth]{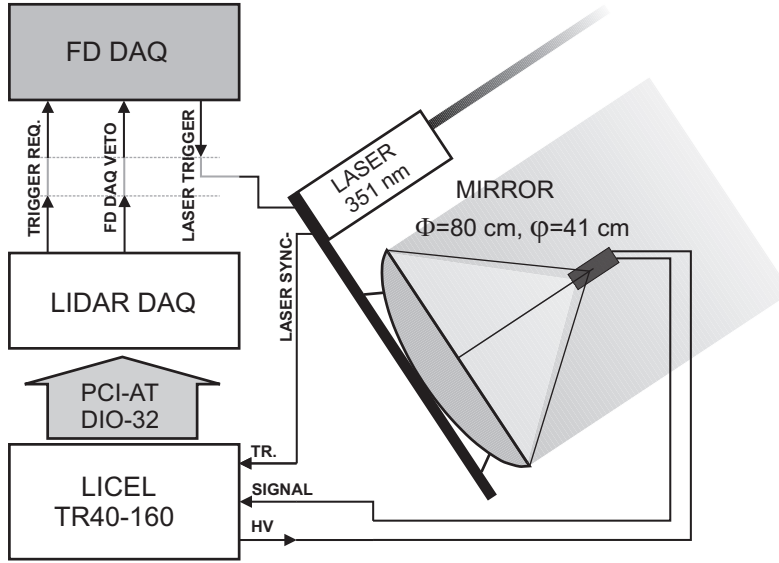}
  \caption{\it Schematic diagram of the Pierre Auger Observatory lidar system.
  Each lidar telescope uses a set of three $\varPhi=80\,\mathrm{cm}$ diameter 
  parabolic mirrors with a focal length of $\varphi=41\,\mathrm{cm}$.}
  \label{f:lidarscheme}
  \vskip1cm
\end{figure}

\begin{figure}[t]
  \centering
  \includegraphics[width=.9\textwidth]{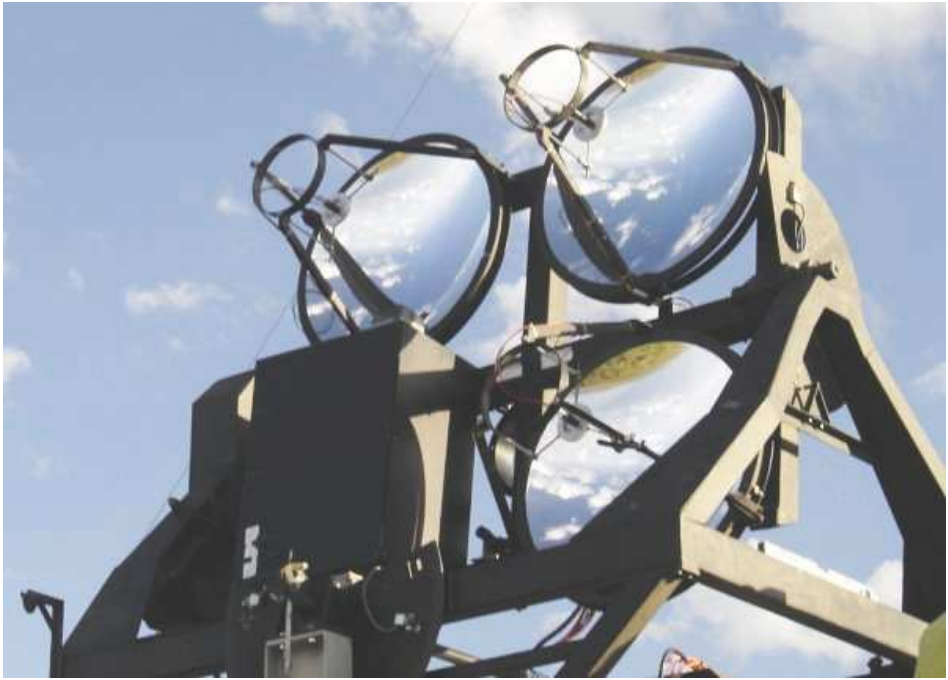}
  \caption{\it The Los Leones lidar system.  Pictured are the 3 mirrors and the box
  which houses the high frequency laser.  The laser shoots from the box in the direction
  of the field of view of the mirrors.}
  \label{f:lidarphoto}
  \vskip1cm
\end{figure}

\section{Lidar Hardware and Data Acquisition}\label{sec:hardware}

As of March 2006, three of the four lidar stations of the Pierre Auger Observatory 
have been fully operational.  The first lidar was installed at the Los Leones site 
in March 2002 and started data taking soon after, mainly to test the impact on FD 
operations and define optimal running conditions.  The lidar at the Coihueco site
began operation in March 2005, and the lidar at the Los Morados site started operation
in March 2006.  The fourth FD at Loma Amarilla is currently under construction and is
expected to be completed by December 2006, with the corresponding lidar station to be 
operational at the same time.  A schematic diagram of a lidar station is shown in
Fig.\,\ref{f:lidarscheme}.  Fig.\,\ref{f:lidarphoto} shows a photograph of the Los
Leones lidar setup.

\subsection{Mount}

Each lidar station has at its core a fully steerable alt-azimuth frame 
built originally for the EAS-TOP experiment~\cite{Aglietta:1993}.  Two DC 
servomotors steer the frame axes with a maximum speed of $2^{\circ}/\mathrm{s}$. 
The absolute pointing direction is known to $0.2^{\circ}$ accuracy.  

The frame is mounted on a $20^{\prime}$ shipping container and is protected from 
the weather by a fully retractable motorized cover during periods when the lidar 
is not operating.
Frame-steering and cover movements are controlled by an MC-204 motion controller
from Control Techniques, which allows the system to be operated both locally 
at the site and remotely via Ethernet.

\subsection{Laser}

Each mount is equipped with a UV laser source and mirrors for the detection of 
the backscattered light.  The choice of a laser for the lidar system is dictated
by the following requirements: the wavelength of the laser has to roughly match the
dominant wavelength of air fluorescence photons; the laser power should be low 
to minimize interference with the FD; and the repetition rate should be high to
reduce data collection time.  To meet these requirements, the lidars are operated 
with diode pumped solid state lasers of type DC30-351 manufactured by Photonics 
Industries.  This laser generates the third harmonic of Nd:YLF at 351\,nm 
and is operated at a repetition rate of 333\,Hz and a per-pulse energy
of roughly $100\,\mathrm{\mu J}$.  The laser wavelength of 351\,nm is at the center
of the nitrogen fluorescence line spectrum, which extends from about 300\,nm to 400\,nm,
with three main spectral lines at 337\,nm, 357\,nm and 391\,nm~\cite{Kakimoto:1996}.

\subsection{Mirrors}

For the collection of the backscattered light, each lidar telescope uses a set 
of three $\varPhi=80\,\mathrm{cm}$ diameter parabolic mirrors with a focal length 
of $\varphi=41\,\mathrm{cm}$ (see Fig.\,\ref{f:lidarscheme}).  The mirrors were
produced using BK7 glass coated with aluminum, and the reflective
surface is protected with $\mathrm{SiO}_2$ coating to ensure the
necessary surface rigidness as well as good UV transmittance. The
average spot size at the focus is $3\,\mathrm{mm}$ FWHM.  Each of the
mirrors is mechanically supported by a Kevlar frame which is in turn
fixed to the telescope frame using a three point system.  This allows 
fine adjustment of the field of view direction to ensure collinearity 
of the mirrors and the laser beam.

\subsection{Photomultiplier and Digitization}

A Hamamatsu R7400U series photomultiplier is used for backscatter light 
detection.  Each mirror has its own photomultiplier, so each lidar telescope 
comprises three independent mirror/photomultiplier systems.  
The photomultiplier reaches a gain of $2\times 10^6$ at the maximum operation 
voltage of $1000\,\mathrm{V}$.  To facilitate the light collection from the 
mirror, the whole active $8\,\mathrm{mm}$-diameter photomultiplier window is 
used.  

Background is suppressed by the means of a broadband UG-1 filter with
$60\%$ transmittance at $353\,\mathrm{nm}$ and FWHM of $50\,\mathrm{nm}$.  
The use of far more selective interference filters is unfortunately not
possible because the extreme speed $f/0.5$ ($\varphi/\varPhi\simeq 0.5$) of
the mirrors leads to a large spread of possible incident angles.  As 
interference filters are very sensitive to the light impact angle, light has 
to hit the filter almost orthogonally, or else the transmitted wavelength can 
shift considerably.  However, one must bear in mind that the lidar is 
constructed to operate during FD data taking, which is only possible on 
moonless nights.  A simple absorption filter is therefore sufficient for 
effective background suppression.

Due to our specific design requirements, a rather long (12\,m) signal
cable between the photomultipliers and digitizers has to be used.  To
minimize the signal dispersion as well as RF interference, UVF-303
series military standard cables are used.  

The signals are digitized
using a Licel TR40-160 three-channel transient recorder.  For analog
detection the signal is amplified and digitized by a 12\,bit 40\,MHz
A/D converter with 16\,k trace length ({\it current mode}).  At the same 
time a fast 250\,MHz discriminator detects single photon events above 
a selected threshold voltage ({\it photon counting mode}).  A combination 
of current and photon counting measurement is used in the subsequent
analysis to increase the dynamic range of the whole system.  The Licel 
recorder is operated using a PC-Linux system through a National Instruments 
digital input-output card (PCI-DIO-32HS) with the Comedi interface within 
the ROOT framework.

\subsection{Trigger}

The lidar is connected to the FD data acquisition (FD DAQ) system by 
means of three optical fibers. Whenever the lidar system wants to 
start a measurement, a trigger request is issued to the FD DAQ.
In response, a logic pulse of frequency $333\,\mathrm{Hz}$ is generated
by the FD GPS clock and transmitted to the laser, which fires 
a single laser pulse for every trigger.  The frequency of $333\,\mathrm{Hz}$ 
corresponds to the maximum acquisition rate of the digitizer for the 
given memory depth ($16\,\mathrm{k}$) and sample rate ($40\,\mathrm{MHz}$). 
The lidar DAQ is triggered by the laser synchronization signal 
generated at every successful laser shot.  Whenever the lidar direction of 
measurement comes close or into the FD field of view, a veto signal which
prevents FD data acquisition can be generated.

The lidar DAQ software is organized in several layers to allow remote or
unattended operation as well as integration into the central Auger DAQ system. 
A run-control program sends the hardware settings and run parameters 
to the DAQ program through a communication server.  The DAQ program 
controls the Licel and photomultiplier settings (tube gain via high voltage
level, photon scaler discriminator level), triggering system, 
telescope steering and cover operation.  Through the DAQ program, the user also
controls the shooting directions and the number of laser shots per shooting angle.
Current and photon counting 
traces are summed for 1000 laser shots in the Licel, stored in a ROOT file 
and sent to the online analysis framework for monitoring.

\begin{figure}[ht]
  \centering
  \includegraphics[width=.9\textwidth]{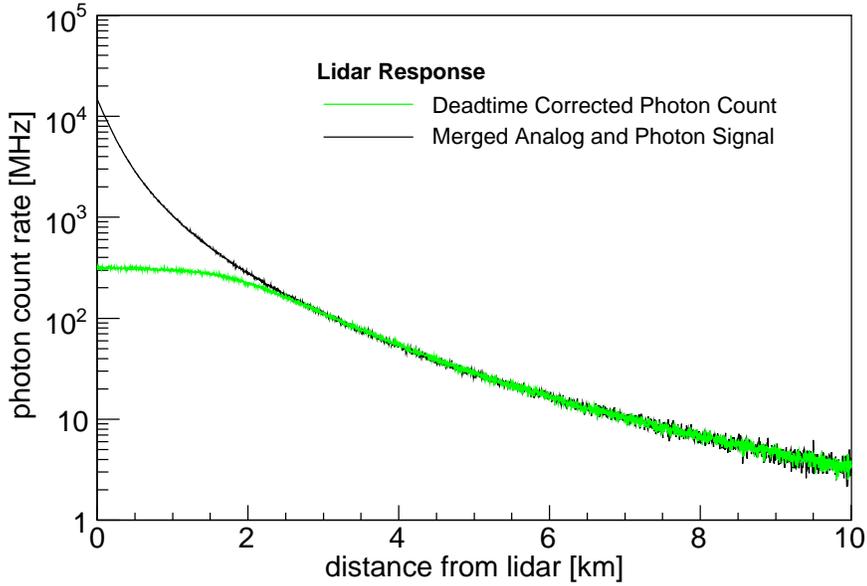}
  \caption{\it Signals from both current and photon counting mode.  
  The figure shows the backscatter signal up to 10\,km distance from the lidar.
  As long as the photon counting trace is saturated, only the current mode trace 
  is used.  When saturation becomes negligible the signal in current mode 
  is fused with the one in photon counting mode.}
  \label{f:signal}
\end{figure}

\begin{figure}[ht]
  \centering
  \includegraphics*[width=0.55\textwidth,angle=0,clip]{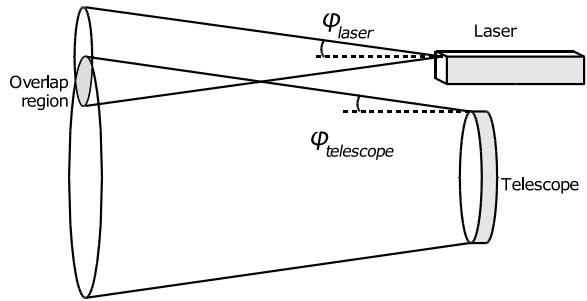}
  \caption{\it A diagram showing the overlap that occurs in our lidar system, where
  the laser beam is emitted parallel and outside the field of view of the telescope.}
  \label{f:overlap} 
  \vskip1cm
\end{figure}

\begin{figure}[t]
  \centering
  \includegraphics*[width=0.7\textwidth,angle=0,clip]{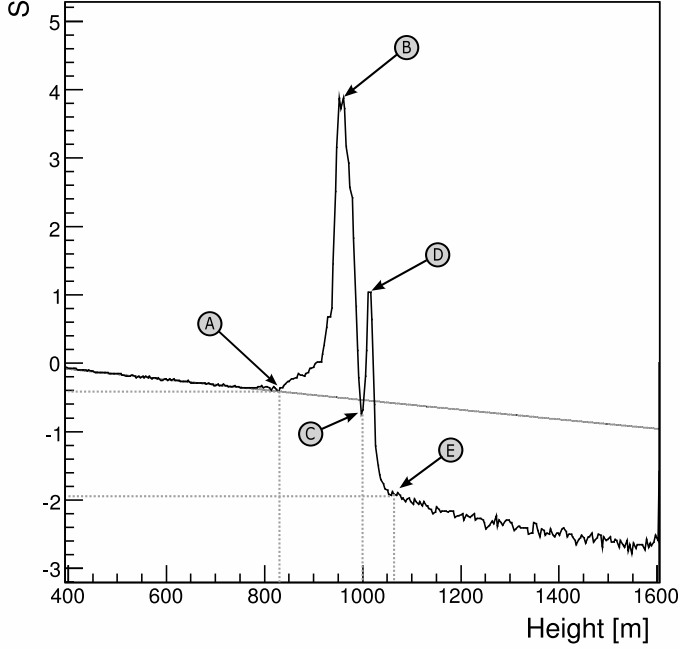}
  \caption{\it The normalized and range-corrected lidar return $S(h)$.  Clouds are seen 
  as strong scattering regions.  The first cloud starts at $A$ and shows the maximum peak 
  at $B$. In this example, a second cloud starts at $C$, shows a peak at $D$ and ends at $E$. 
  The grey curve is a simulated lidar response for a purely molecular atmosphere.}
  \label{f:scloud}
  \vskip1cm
\end{figure}

\subsection{Signal Treatment}\label{subsec:signaltreatment}

As mentioned above, the Licel module records backscattered light measurements 
in current mode and photon counting mode.  Current mode maximizes the near-field 
spatial resolution, but it is suitable only for the first few kilometers where the 
signal is sufficiently high.  As the signal decreases as the square of the distance, 
photon counting is required in order to obtain information about the atmosphere 
at large distances.  On the other hand, the photon counting saturates in the near-field
due to limitations of the Licel.  At distances less than 5\,km from the lidar station, 
the light level causes a rate greater than 10\,MHz and the deadtime starts to be an 
issue ($\geq 5\,\%)$.  However, the combination of current and photon counting mode 
covers the full dynamic range of the return signal from near the detector out to a 
distance of 25\,km.  Fig.\,\ref{f:signal} shows an example for a signal in current 
and photon counting mode.

In order to combine the current and photon counting signals, the ratio of the two 
signals is calculated.  A range in which this ratio is almost constant is identified,
usually when the photon count rate is under 10\,MHz, and signals are merged in that 
region.  In the first kilometers this condition is not met due to the fact that the 
photon counting mode saturates for high photon rates (see Fig.\,\ref{f:signal}).  
Typically, the optimal merging region is 5-10\,km from the detector, where both the 
current and photon counting signals are valid.

The derived signal can be parameterized by the so-called lidar equation:

\begin{equation}
	P(r)=P_0\frac{ct_0}{2}\beta(r)\frac{A}{r^2}e^{-2\tau(r)}
            =P_0\frac{ct_0}{2}\beta(r)\frac{A}{r^2}
             e^{-2\int_0^r\alpha(r')dr'} \hspace{10pt},
	\label{eqlidar}
\end{equation}

where $P(r)$ is the signal received at time $t$ from photons scattered at a distance 
$r$ from the lidar, $P_0$ is the transmitted laser power, $t_0$ is the laser pulse duration,
$\beta(r)$ is the backscattering coefficient, $\tau(r)$ is the optical depth,
$\alpha(r)$ is the extinction coefficient, and $A$ is the effective receiving 
area of the detector. $A$ is proportional to the overlap of the telescope field 
of view with the laser beam (shown in Fig.\,\ref{f:overlap}).  
Over the range of distances where the laser beam and mirror viewing field only
partially overlap, it is possible to experimentally determine an overlap function
from horizontal scans~\cite{Sasano:1979}.
However, since we restrict our analysis to a range starting at $r_{0}=750$\,m, 
where the overlap is complete, we do not apply an overlap correction.

The far limit $r_{f}$ of our range is defined as the distance at which the 
signal is negligible as compared to the background, typically 25\,km.
In the region from $r_0$ to $r_f$, it is convenient to express the return signal 
as a function of height $h$ in terms of a range-corrected and normalized auxiliary 
function, $S(h)$:

\begin{equation}
	S(h)=\ln \frac{P(h)h^2}{P(h_n)h_n^2}
            =\ln{\frac{\beta(h)}{\beta(h_n)}}-2\tau(h;h_n)\sec(\theta) \hspace{10pt}.
	\label{eqS}
\end{equation}

In this equation, $P(h)$ is the signal at height $h$, $\tau(h;h_n)$ is the optical 
depth calculated in the range $[h_n,h]$, and $\theta$ is the lidar inclination angle 
from the zenith.  The normalization height $h_n$ is a fixed height to normalize $P$,
chosen such that at $h_n$, the entire signal is in the field of view of the mirrors.

Clouds are visible as strong localized scattering sources.  Fig.\,\ref{f:scloud} 
shows an example of $S(h)$ in the presence of a cloud layer starting at a height of 
about 830\,m.  Also shown is the simulated lidar response for a purely molecular
atmosphere.  The figure illustrates that it is possible to resolve several layers 
of clouds.  A second cloud layer can be seen at an altitude of about 1000\,m.
The cloud finding algorithm applied in this analysis is described in detail in 
Section~\ref{subsec:anclouds}.

\section{Operation}\label{sec:ops}

\subsection{Current Status}\label{subsec:opsstatus}

The lidar stations were designed to be operated remotely from the observatory's 
central campus in Malarg\"ue.  There, a computer is used to centralize the 
operation and issue all the startup commands to the three existing lidar stations
and also to monitor the quality of the data being collected.

The remote operation of systems with this level of complexity presents a number 
of challenges.  In order to achieve a safe handling of the telescopes, various
software routines and hardware devices have been installed to monitor the
performance and status of lidar operations.  These monitoring subsystems include 
programs used to collect weather related information (mainly rain and
wind speed data).  The presence of ambient light and the status of the power 
supply are monitored as well.  In the occurrence of an external event such as 
rain that could jeopardize the lidar equipment, these subsystems assume control 
of the station by parking the telescope and closing the cover.

\begin{figure}[t]
  \centering
  \includegraphics[width=.9\textwidth]{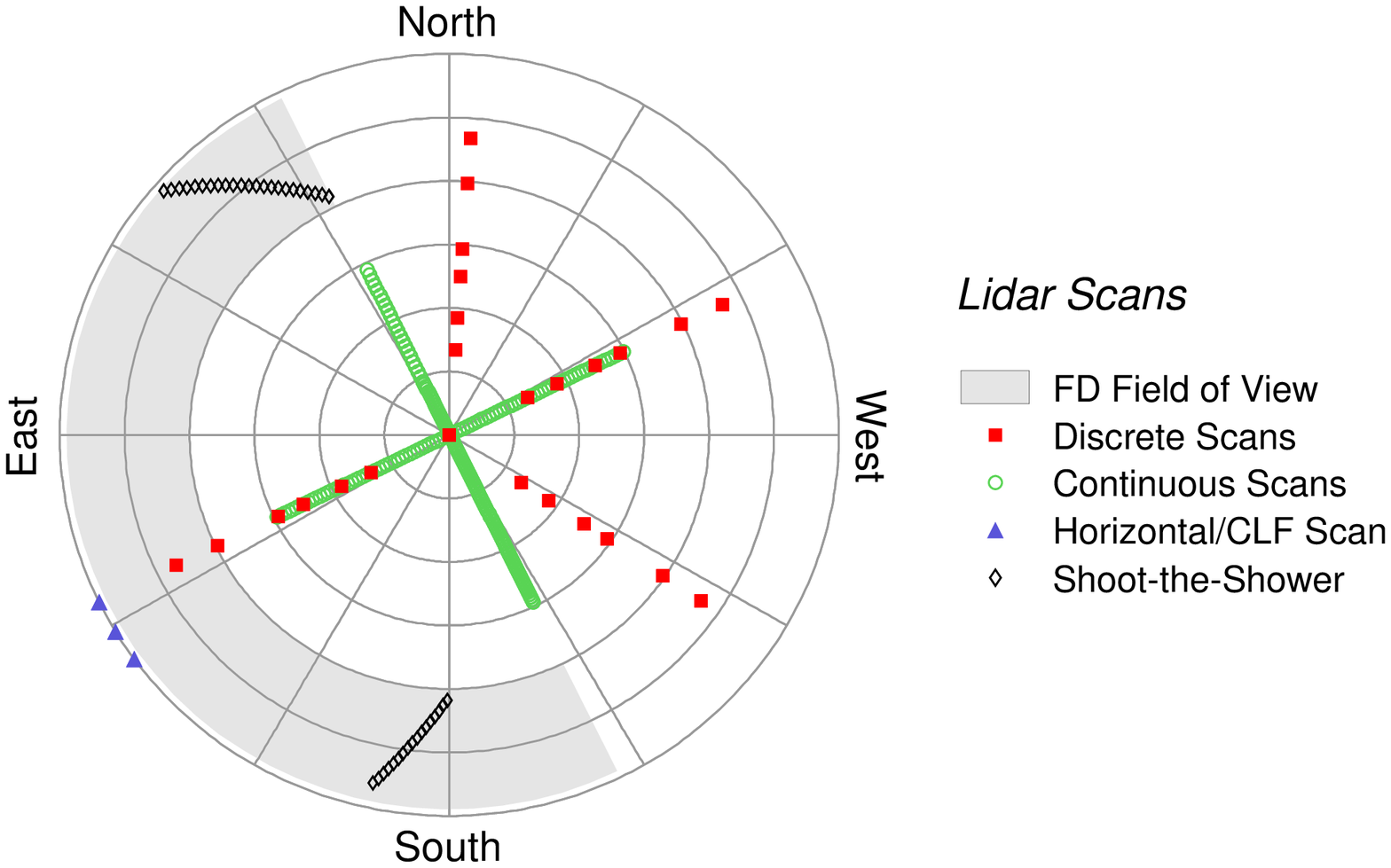}
  \caption{\it A typical night of lidar shooting activities at Coihueco,
  shown in an azimuthal equatorial projection of the sky.  Depicted are
  the coordinates for the lidar automatic shooting strategy, which
  comprises: discrete sweeps for atmospheric parameter estimation;
  continuous sweeps for cloud detection; horizontal shots toward the
  Central Laser Facility (CLF) for calibration; and shoot-the-shower scans
  to probe the tracks of important showers viewed by the fluorescence
  detector (FD).  The Coihueco FD field of view is shown in gray.}
  \label{f:scandir}
  \vskip1cm
\end{figure}

\begin{figure}[t]
  \centering
  \includegraphics[width=1.\textwidth]{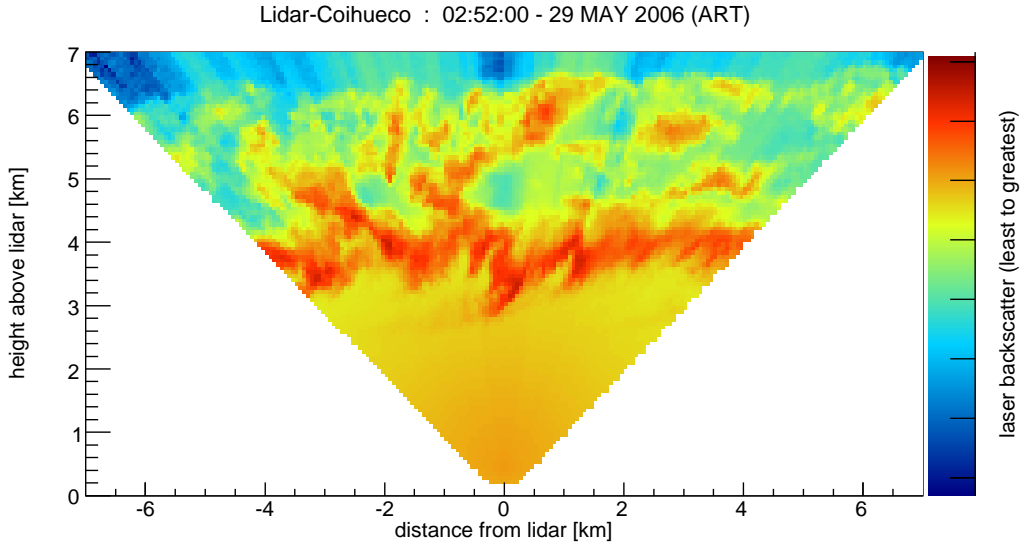}
  \caption{\it Result of a typical continuous lidar scan.  Shown is the intensity of 
  backscattered light as a function of height and horizontal distance to the lidar 
  station at (0,0).  A cloud layer around 3.5\,km height is clearly visible in this scan.} 
  \label{f:contscan}
  \vskip1cm
\end{figure}

\subsection{Typical Operation}\label{subsec:opstypical}

Lidar operation starts at astronomical twilight.
After the telescope cover is opened, an initialization procedure is executed 
to calibrate the incremental encoders used to determine the telescope position.
  
A webcam located in the interior of the telescope cover is used to confirm visually that
these tasks are executed correctly. In this way, before starting a run, the operator
has information about the status of the telescope in real time and about the weather
conditions at each site through the information being sent to the lidar web 
site.

Following initialization, the system enters an operational mode called \emph{AutoScan}. 
In AutoScan mode, the telescope performs a cycle of steering scripts unless otherwise 
interrupted until the end of the night.  When the laser is fired, the telescope position 
is determined by the coordinates contained in these scripts.  There are four main steering 
strategies: three making up the AutoScan pattern and a fourth, shoot-the-shower, that
periodically interrupts the AutoScan.  These strategies are discussed below, and
Fig.\,\ref{f:scandir} shows, in an azimuthal equatorial projection of the sky,
the firing pattern for a typical night of lidar activities at Coihueco.

\begin{itemize}
    \item \emph{Continuous scans:} In this scan, the telescope is moved between 
    two extreme positions with a fixed angular speed while the laser is fired.
    The telescope sweeps the sky along two orthogonal paths with fixed azimuthal 
    angle, one of which is along the central FD azimuth ($90^{\circ}$).  Along 
    both paths, the maximum zenith angle is $45^{\circ}$.  The continuous sweeps 
    are constrained to take 10 minutes per path from start to finish.  Along each 
    path, the lidar performs on the order of 100 measurements with 1000 shots per 
    measurement.  The purpose of these scans is to provide useful data for simple 
    cloud detection techniques (see Section~\ref{subsec:anclouds} for details)  
    and to probe the atmosphere for horizontal homogeneity.  An example of the data 
    produced by this kind of scan is shown in Fig.\,\ref{f:contscan}.  
    \item \emph{Discrete scans:} In this scan, the telescope is positioned at a set 
    of particular coordinates to accumulate larger statistics at a few locations.
    As indicated in Fig.\,\ref{f:scandir}, these measurements are performed at 
    6 discrete zenith angles for 4 different azimuth angles, and directly overhead
    (zenith angle $0^{\circ}$).  To accumulate large statistics, 12 measurements 
    are performed at each location.  Each measurement consists of 1000 laser shots 
    run at 333\,Hz.  The combined duration of the two discrete sweeps is about 
    30 minutes.  The data obtained in this mode are useful to determine the 
    vertical distribution of aerosols in the atmosphere.
    \item \emph{Horizontal and CLF shots:} In this mode, the laser fires horizontally
    in 3 different directions towards the location of the CLF.  Three measurements 
    with 1000 shots per measurement are performed.  The data collected in this scanning mode 
    are used to detect low-lying aerosols and also to determine the horizontal 
    attenuation between the CLF and the FD telescopes for comparison with measurements 
    made by other atmospheric monitoring systems.  The total duration is about 3 minutes.
    \item \emph{Shoot-the-Shower (StS):} This rapid response mode is used to measure the 
    atmospheric attenuation in the line of sight between the FD telescopes and a detected
    cosmic ray shower. This scanning mode suspends any of the previously mentioned sweeps.
    It will be described in further detail in section \ref{subsec:opstypical}.
\end{itemize}

A complete scanning cycle, excluding StS, takes about 60 minutes to complete.  All scans 
are therefore performed on an hourly basis.  The maximum length of a lidar running night
depends on the length of astronomical twilight and varies over the course of the year
from less than five hours during the summer to almost fourteen hours during the winter.  

As shown in Fig.\,\ref{f:scandir}, some shooting positions are very close to or inside the 
field of view of the FD telescopes. In order to prevent the detection of a large number of
spurious FD events generated by the lidar shooting activity, buffer zones have been delimited 
around the FD fields of view.  Every time the laser is fired inside this buffer zone, the 
FD DAQ is inhibited in order to avoid any interference.  This is accomplished by sending a veto 
signal from the lidar to the FD when the laser is ready to fire.  The total FD deadtime
introduced by all lidar operations is less than $2\,\%$.

\begin{figure}[t]
  \centering
  \includegraphics*[width=0.55\textwidth,angle=0,clip]{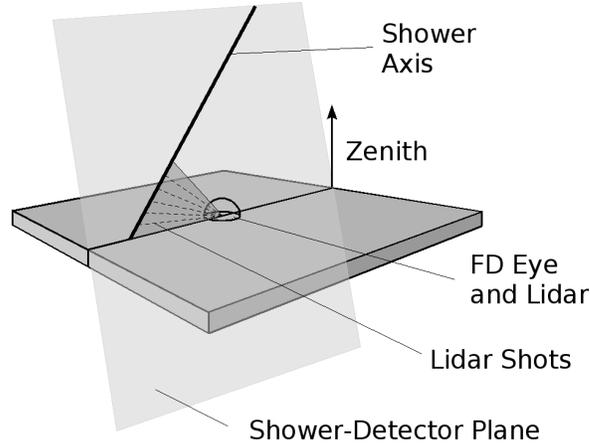}
  \caption{\it The geometry of shoot-the-shower (StS).  The lidar station 
  at a fluorescence detector site initiates shots in the shower-detector
  plane within the zenith field of view of the FDs (approximately 
  $\it 0^{\circ}$ to $\it 30^{\circ}$ in elevation).  Up to 60 pointing 
  directions, with 1000 laser shots per pointing, are allowed per StS.}
  \label{f:lidar-sts}
  \vskip1cm
\end{figure}

\begin{figure}[t]
  \centering
  \includegraphics*[width=1.0\textwidth,angle=0,clip]{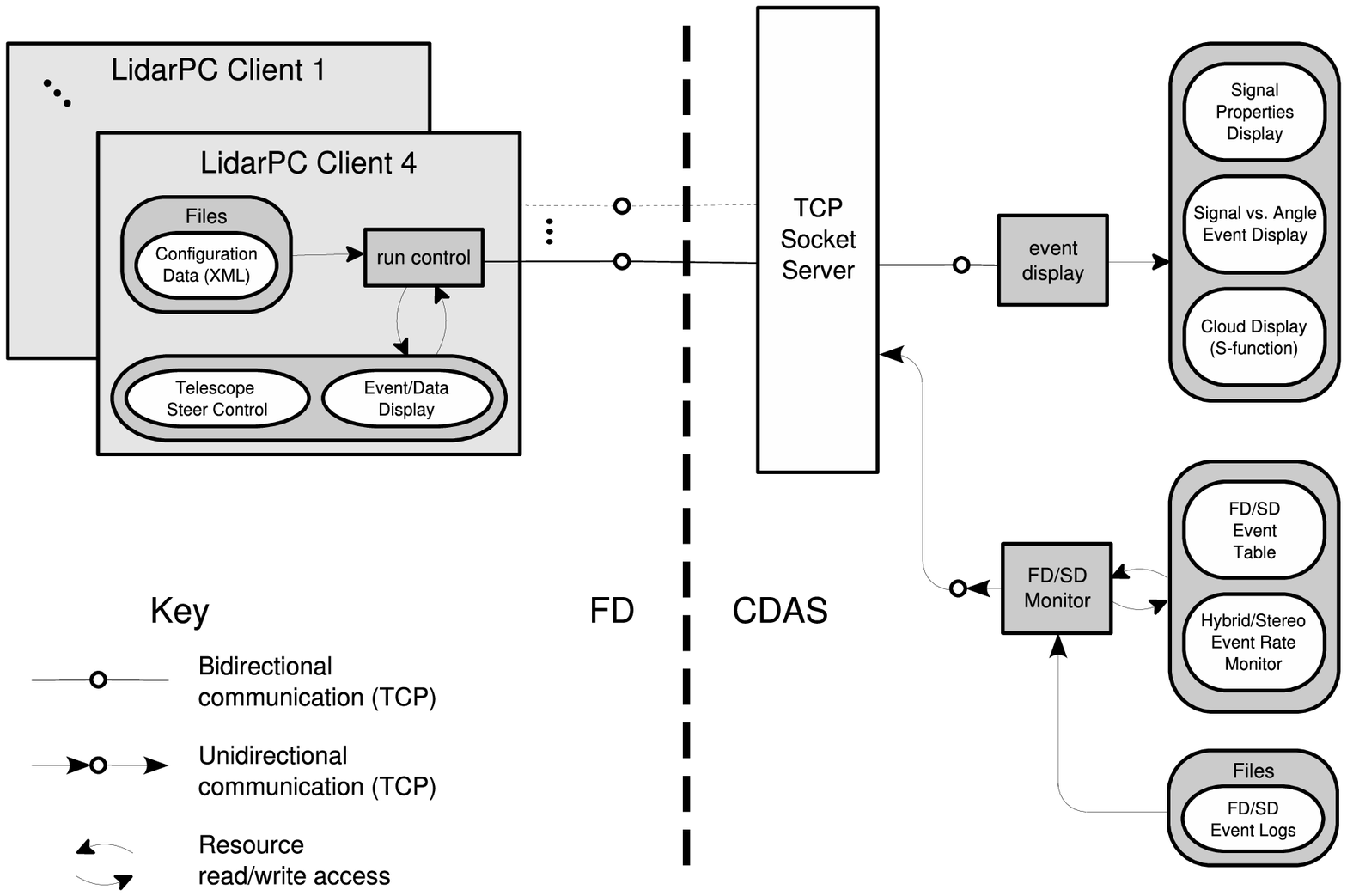}
  \caption{\it Communication between the Pierre Auger Observatory central data 
           acquisition system (CDAS)
           and the independent lidar stations running in each FD subnetwork.
           A data monitoring client in CDAS checks for hybrid and stereo 
           events; via TCP, it sends StS requests to the LidarPCs, where the 
           lidar run control program executes shoot-the-shower.}
  \label{f:lidar-soft}
  \vskip1cm
\end{figure}

\subsection{Shoot-the-Shower}\label{subsec:opssts}

A primary design requirement of the lidar system is that it probe the
atmosphere along the tracks of cosmic rays observed by the FDs.  This function,
called shoot-the-shower (StS), exists to recognize unusual and highly localized 
atmospheric conditions in the vicinity of individual air showers of high interest.
Conditions that can affect FD observations at different times of the year
include the presence of low and fast clouds, and low-level aerosols due
to fog, dust, or land fires.

The basic operation of StS is depicted in Fig.\,\ref{f:lidar-sts}.  The axis of
a cosmic ray air shower, when projected onto the field of view of an observing
air fluorescence detector, defines a plane called the shower-detector plane, or
SDP.  When a lidar station shoots the shower, it performs a series of laser
shots within this plane, determining the atmospheric transmission between the
shower segment and the FD.  For a given shower, up to 60 pointing directions
with 1000 laser shots per pointing are allowed, all within the FD field of view.

To operate effectively, the lidar must shoot showers of importance
with a minimum of delay after each shower occurs and
with minimum disruption to the normal operation of the FD.  The showers of
primary interest for StS are hybrid events involving the FDs and the SD, since
these set the energy scale of the observatory.  FD stereo events, which are of
high energy and typically at large distances from the FDs, are also good StS
candidates.

Physics events are made available to the lidar by picking off the SD and FD
data stream sent continuously to the observatory's central data acquisition
system (CDAS), located on the central campus.  A program running 
in CDAS monitors incoming data, checks for hybrid and stereo events, and 
calculates StS trajectories.  StS requests are sent to the lidar station data 
acquisition and control PCs (LidarPCs), which initiate the laser shots.

Communication between the lidar stations and CDAS, shown schematically in
Fig.\,\ref{f:lidar-soft}, occurs over the network links between the central
campus and the four FD subnetworks.  Each LidarPC listens on an open TCP port
for StS requests from the FD/SD monitor running in CDAS.  As the lidar shoots,
it returns basic diagnostic information, such as shooting directions and PMT
signal peaks, back to CDAS for display.  A TCP socket server operating in CDAS
manages the bidirectional flow of data packets between the central campus and
the lidar stations.

Lidars receiving StS requests from CDAS automatically stop the default shooting
operation (AutoScan), move to the FD field of view, initiate the StS, and then
resume the AutoScan when StS is complete.  If the lidar receives an StS request
while shooting another shower, the request is pushed into a queue for later
processing.  

To minimize the FD dead-time introduced by StS, the lidar software cuts on
the number of SD tanks participating in the event, which is a rough measure
of the primary cosmic ray energy.  This cut reduces the number of requests to 
several per FD site per night.  A further reduction in the rate is achieved by 
rejecting events caused by known artificial light sources in the detector, such 
as other lidar stations and the CLF.  In addition, the intensity of the 
high-repetition laser means that the lidar must carefully avoid incidents of 
cross fire into other unvetoed FDs during StS.  Therefore, angular windows 
in azimuth and in zenith are defined around each FD; the lidar is
forbidden from entering these windows during StS.  

The operation of StS in the field of view of the photomultiplier cameras raises 
the issue of possible long term effects on the phototubes themselves.  Although 
FD data acquisition is inhibited during StS, the photomultipliers continue to
operate at high voltage during their exposure to powerful nearby laser shots.  
However, since the shooting rate is one to two shots per FD per night, the effect 
of the StS is not significant in comparison to other strong and persistent
light sources.  These sources include the typical night sky background
with its large number of bright stars, and heavy lightning activity
during the summer months. In addition, a comparison of tube noise directly 
before and after StS events shows no significant effect of the shooting activity
on the tube noise.

\begin{figure}[t]
  \centering
  \includegraphics*[width=0.7\textwidth,angle=0,clip]{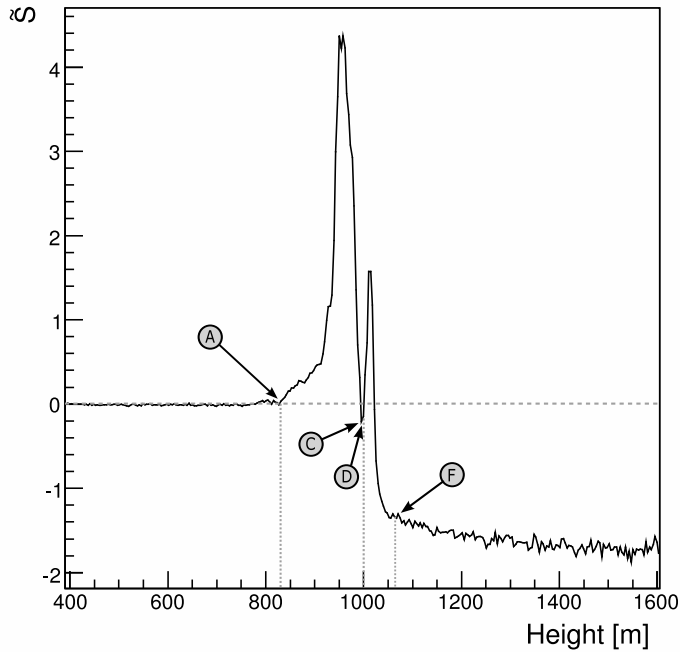}
  \caption{\it The normalized and range-corrected lidar return $S(h)$ after subtraction 
  of a simulated return that assumes a purely molecular atmosphere.}
  \label{f:smolcloud}
  \vskip1cm
\end{figure}

\begin{figure}[t]
  \centering
  \includegraphics*[width=0.6\textwidth,angle=270,clip]{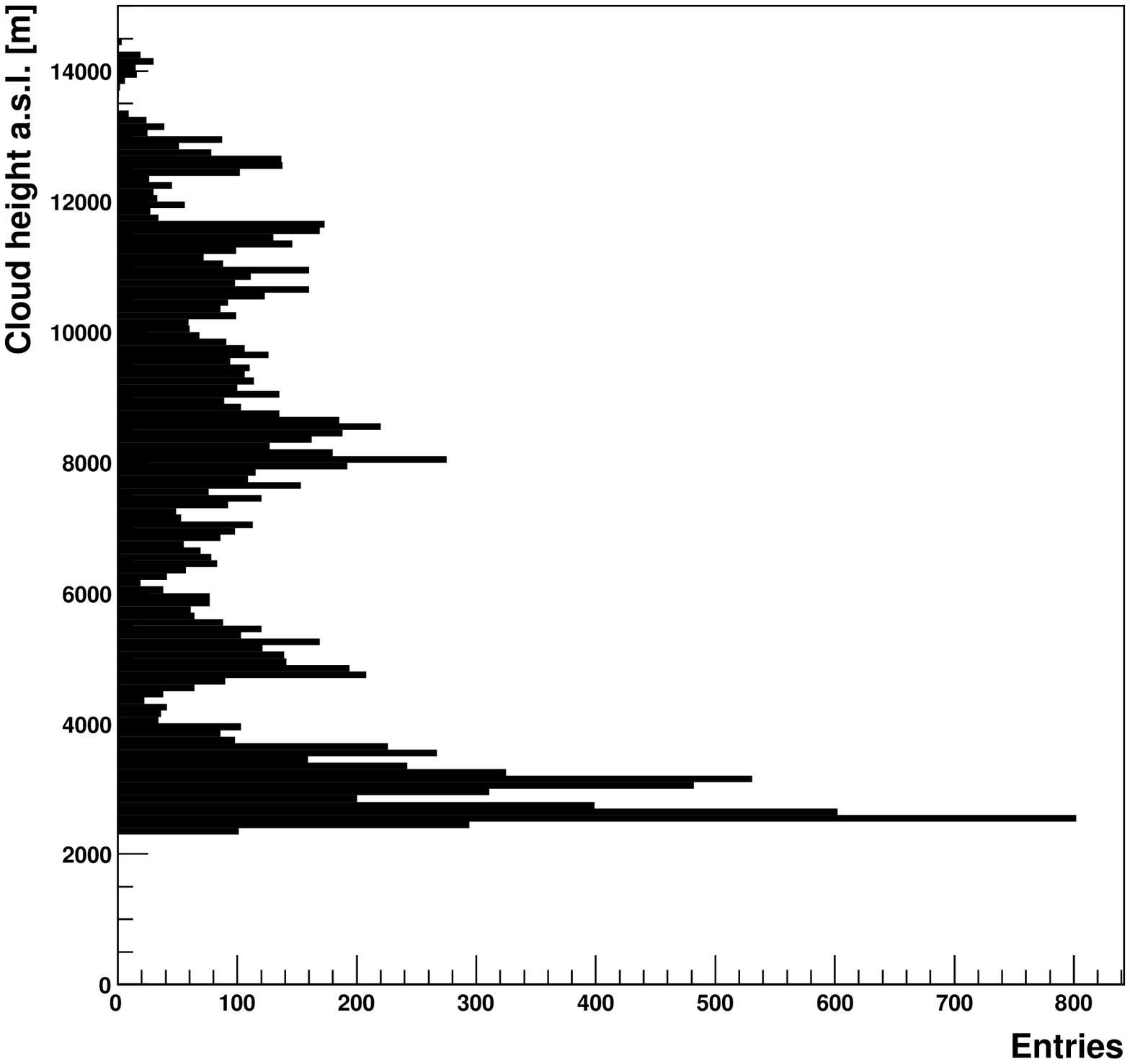}
  \includegraphics*[width=0.6\textwidth,angle=270,clip]{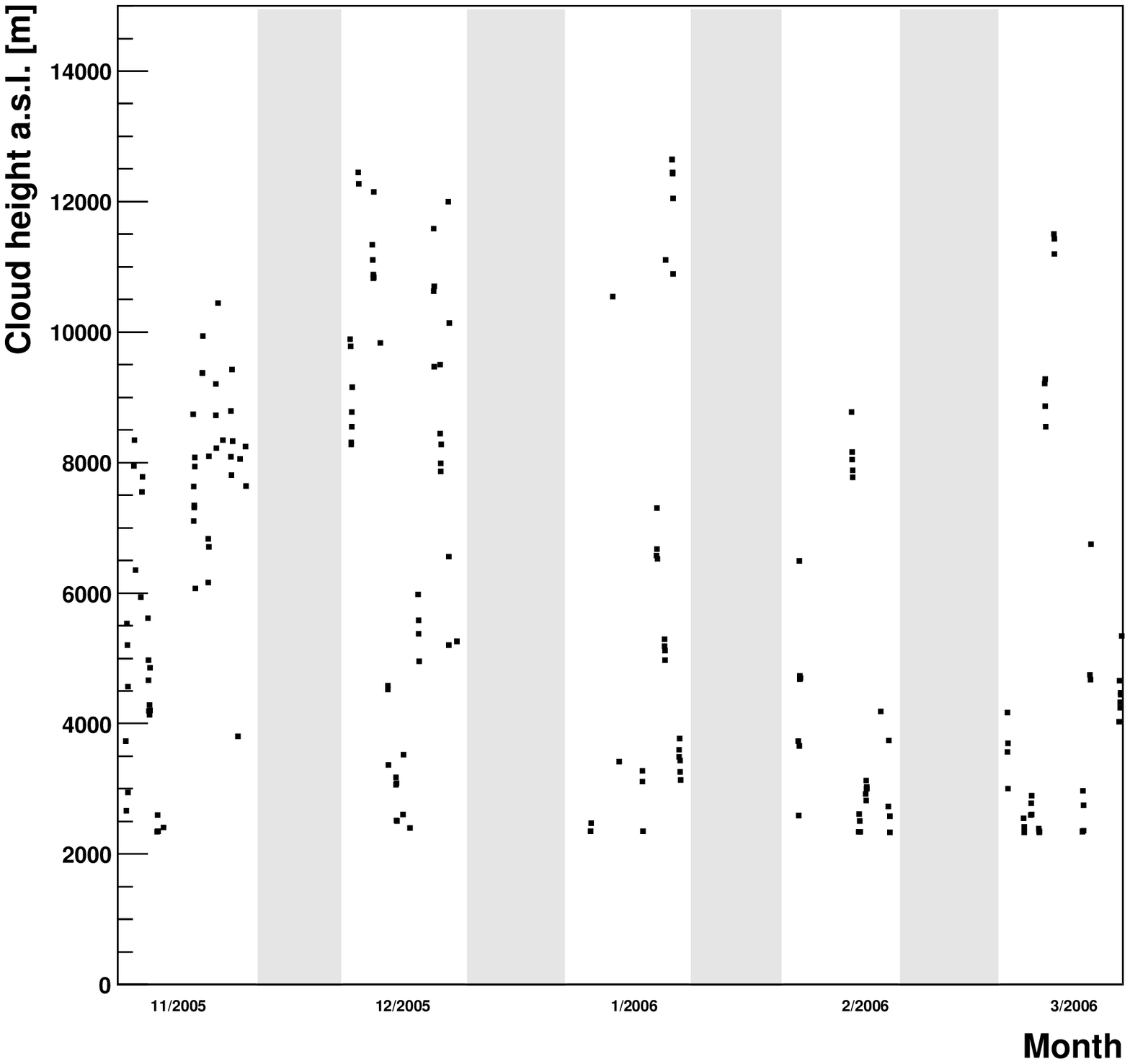}
  \caption{\it Distribution of the height above sea level (a.s.l.) of the lowest cloud 
  layer (upper plot), and height of the lowest cloud layer as a function of time 
  (lower plot) for data taken between 24 October 2005 and 7 March 2006 with 
  the Coihueco lidar.  The grey areas in the lower plot indicate the periods 
  when the moon is too bright for FD data taking.  The plots have 1 entry for 
  each hour when a cloud was detected.}
  \label{f:allheights}
\end{figure}

\begin{figure}[ht]
  \centering
  \includegraphics*[width=0.6\textwidth,angle=270,clip]{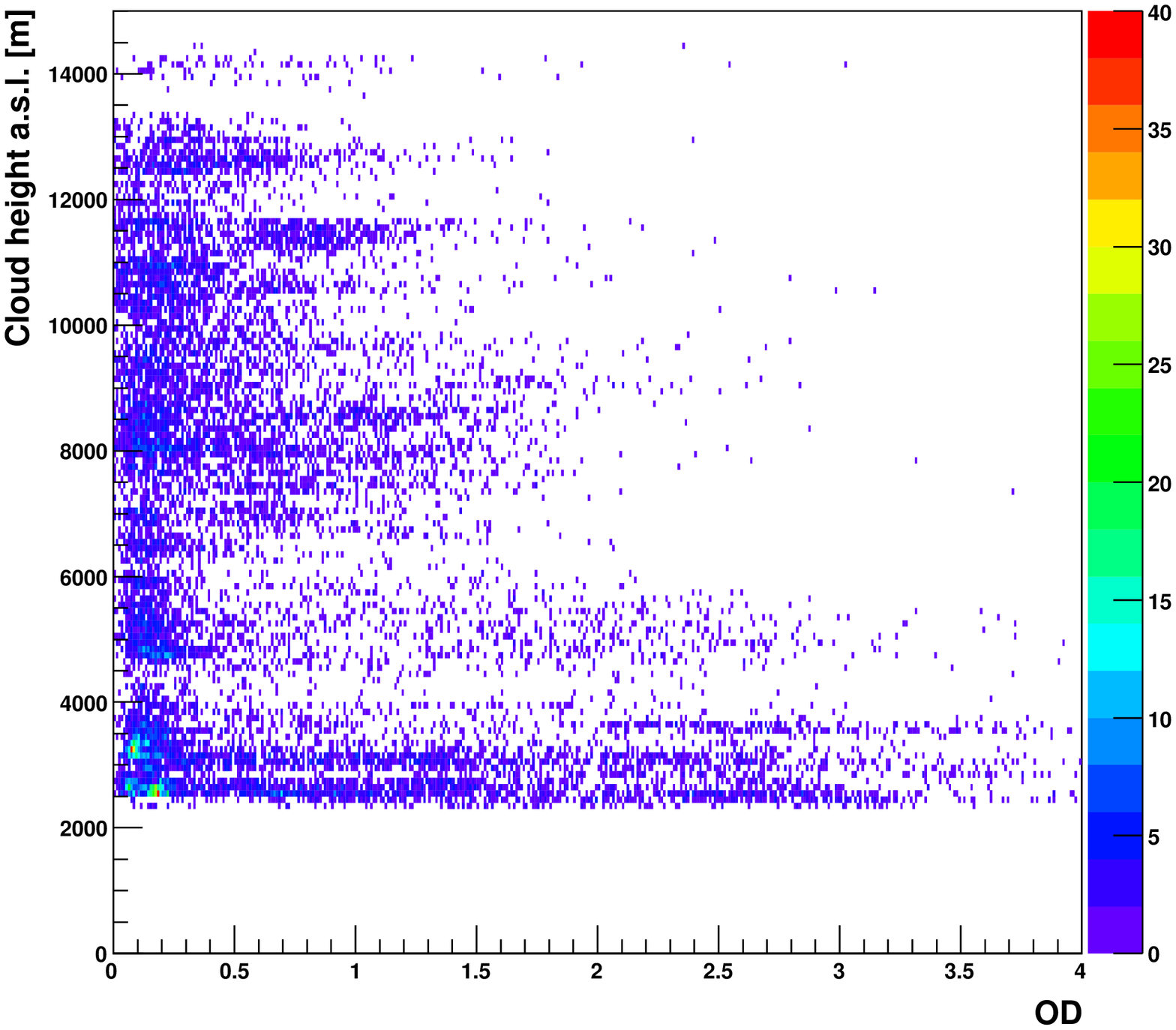}
  \caption{\it Height above sea level (a.s.l.) of the lowest cloud layer versus 
  vertical aerorosal optical depth inside the lowest cloud layer observed by the
  Coihueco lidar between 24 October 2005 and 7 March 2006.}
  \label{f:odvsheight}
  \vskip1cm
\end{figure}

\section{Analysis}\label{sec:analysis}

In standard operation mode, the two goals of the lidar system are (1) to provide 
hourly measurements of vertical aerosol optical depth and backscatter coefficient 
as a function of height at each FD site, and (2) to provide hourly information on 
cloud coverage, cloud height, and the optical depth of cloud layers.  The analysis 
of lidar scans to extract the vertical aerosol optical depth is discussed in detail 
in~\cite{Veberic:2002}.  Here, we will describe the analysis of cloud coverage and
optical depths of cloud layers.  
These measurements are particularly important for the accurate estimation 
of the FD detection aperture.  While they do not affect the SD aperture,
clouds within the fields of view of the FDs distort the light profile
of showers and can compromise the energy calibration.

\subsection{Cloud Detection}\label{subsec:anclouds}

As described in Section\,\ref{subsec:signaltreatment}, clouds are detected as
strong localized scattering sources, and the timing of the backscattered light 
gives the distance from the cloud to the lidar.  The return signal is expressed 
in terms of the $S$ function (Eq.\,\ref{eqS}).

To automatically detect clouds, it is useful to calculate a new function 
$\tilde{S}(h)$ by subtracting the $S$ function for a simulated purely molecular 
atmosphere from the measured $S$ function:

\begin{eqnarray}	
       \tilde{S}(h) & = & S-S_{mol} \nonumber \\
                    & = & S-\ln{\left[\frac{\beta_{mol}(h)}{\beta_{mol}(h_n)}\right]}
                          +2\tau_{mol}(h;h_n)\sec(\theta) \nonumber \\
                     & = & \ln{\left[\frac{\beta(h)}{\beta(h_n)}
                          \frac{\beta_{mol}(h_n)}{\beta_{mol}(h)}\right]}
                          -2\tau_{aer}(h;h_n)\sec(\theta)  \hspace{10pt},
\label{eqSsub}
\end{eqnarray}

where $h_n$ is the normalization height (as in Eq.\,\ref{eqS}), $\beta(h)$ is the 
backscattering coefficient, $\tau(h;h_n)$ is the optical depth calculated in the 
range $[h_n,h]$, and $\theta$ is the lidar inclination angle from the zenith.

The properties of the molecular atmosphere at the Pierre Auger Observatory site 
are very well-known thanks to an extensive balloon-launching program which has
produced a detailed database including temperature, pressure, and density profiles
over the site~\cite{Bluemer:2005}.  Data from these balloon flights, typically 
performed every 5 days, are used to create monthly models of the molecular atmosphere 
at the Malarg\"ue site.  In the lidar analysis we use these monthly models for the 
calculation of $\tilde{S}(h)$.  By using the monthly models, seasonal variations 
in the molecular atmosphere are accounted for, but day-to-day variations, which 
can be substantial~\cite{Wilczynska:2005}, are not.  These short term variations
do not affect the cloud finding algorithm, and their effect on the calculation of 
the optical depth of the cloud layer is negligible.
As shown in Fig.\,\ref{f:smolcloud}, $\tilde{S}(h)$ is approximately
constant before the cloud, and has a non-zero slope inside the cloud.  We 
apply a second-derivative method~\cite{Tonachini:2007} to identify cloud 
candidates and obtain the cloud thickness.

In cases where several cloud layers are present, the influence of partial 
overlaps between cloud layers causes an inaccurate estimation of the optical depth of
the first cloud.  For this reason, very near clouds separated by less than 10\,m are 
treated as a single cloud layer, and the optical depth is calculated only for the 
combination of the layers.

In order to reduce the possibility of a spurious cloud detection, we take advantage
of the fact that each lidar has three individual mirrors and photomultipliers, 
so the same signal is detected independently several times.  To discriminate against
noise, we require a cloud to be detected in multiple mirrors.  
Signals from different mirrors of the same lidar are therefore compared before
a cloud is stored and a cloud detection is reported.

\subsection{Optical Depth of the Cloud Layer}

The application of the cloud detection algorithm to lidar scans can provide 
a variety of useful data.  Collecting all the scans during an hour, the lowest 
layer of clouds is identified and the proportional sky coverage is calculated.
Fig.\,\ref{f:allheights} shows the distribution of the height of the lowest 
detected cloud layer (upper plot) and the height of the lowest layer as a function
of time (lower plot) for data taken between 24 October 2005 and 7 March 2006  
with the Coihueco lidar.  Typical cloud layers can be identified, in particular
a recurrent cloud layer between 2000\,m and 3000\,m above sea level.

Once the lowest clouds have been identified, their effect on the light propagation 
is estimated.  The turbidity of a layer of thickness $H$ can be described by 
a transmission factor

\begin{equation}
	T(H)=e^{-\tau(H;0)}  \hspace{10pt},
\label{eqTr}
\end{equation}

where $\tau(H;0)$ is the total optical depth. 

Consider, for instance, a cloud at a height $h_A$ that ends at 
a height $h_F$.  From a lidar scan, the auxiliary function $S(h)$, given 
by Eq.\,\ref{eqS}, is obtained, and $\tilde{S}(h)$ is calculated by using Eq.\,\ref{eqSsub}.
The difference between the values assumed by $\tilde{S}(h)$ in $h_A$ and $h_F$ is:

\begin{equation}
	\Delta\tilde{S}(h_F;h_A)=\ln{\left[\frac{\beta(h_F)}
					{\beta_{mol}(h_F)}
					\frac{\beta_{mol}(h_A)}{\beta(h_A)}\right]}
					-2\tau_{aer}(h_F;h_A)\sec{\theta}  \hspace{10pt},
\label{eqSdiff}
\end{equation}

where $\tau_{aer}(h_F;h_A)$ is due to scattering and absorption of the light by the cloud.  

At heights greater than 2\,km above ground level a quasi-molecular atmosphere can be assumed
in the proximity of clouds.  Therefore, $\beta\simeq\beta_{mol}$ at $h_A$ and $h_F$, and 
Eq. \ref{eqSdiff} becomes:

\begin{equation}
	\Delta\tilde{S}(h_F;h_A)\simeq-2\tau_{aer}(h_F;h_A)\sec{\theta}  \hspace{10pt}.
\label{eqSdiffsimpl}
\end{equation}

In this way, the cloud optical depth can be estimated directly from the signal:

\begin{equation}
	\tau_{aer}(h_F;h_A)\simeq-\frac{1}{2}\Delta\tilde{S}(h_F;h_A)\cos(\theta).
	\label{tau}
\end{equation}

Using this information, the mean optical depth inside the lowest cloud layer 
is calculated for every hour of FD data taking.  As an example of the capability 
of the lidar system to detect and characterize clouds, Fig.\,\ref{f:odvsheight} 
shows the height of the lowest cloud layer versus the optical depth inside the 
lowest cloud layer observed with the Coihueco lidar between 24 October 2005 
and 7 March 2006.  The plot indicates that low altitude clouds tend to cover a 
wider range of optical depths than clouds at higher altitude.

The atmospheric monitoring facilities of the Pierre Auger Observatory also include
customized infrared cloud cameras at each FD site~\cite{Cester:2005}.  These cameras 
scan the entire sky every 15 minutes.  The pictures are processed and used to fill 
a database with information on the cloud coverage.  To first order, the images from 
the cloud camera provide a crude cross-check of the efficiency of the lidar cloud 
finding algorithm.  In practice, the two cloud analyses are complementary, with the 
lidar providing the height and optical properties of cloud layers and the cloud 
cameras providing a complete two-dimensional image of the cloud coverage.

\begin{figure}[ht]
  \centering
  \rotatebox{270}{\includegraphics[width=.6\textwidth]{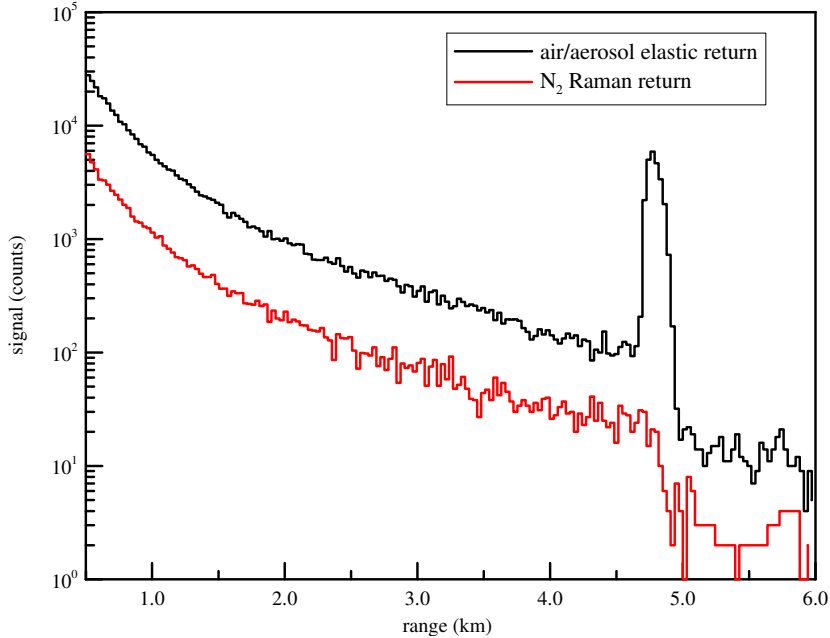}}
  \caption{\it Comparison of the lidar returns in the elastic and nitrogen channels.
  A cloud layer at about 4.8\,km height appears as an attenuation in the nitrogen channel,
  but as an increase in the signal in the elastic channel due to the enhanced backscattering.}
  \label{f:raman}
  \vskip1cm
\end{figure}

\section{Inelastic Raman Lidar}\label{sec:future}

\subsection{Principle of Operation}

In addition to operating the elastic backscatter lidars, the Pierre Auger 
collaboration is currently also evaluating the usefulness of a Raman lidar 
operated in conjunction with the elastic lidar at Los Leones.

The Raman technique is based on inelastic Raman scattering, a secondary component of 
molecular scattering in the atmosphere.  The inelastic component is suppressed 
compared to elastic Rayleigh scattering since the Raman scattering cross section 
is about three orders of magnitude smaller than the corresponding Rayleigh cross 
section.  In Raman scattering, the scattered photon suffers a frequency shift that 
is characteristic of the stationary energy states of the irradiated molecule.  
With Raman spectroscopy, it is therefore possible to identify and quantify 
traces of molecules in a gas mixture.

Raman lidars operate on the same principle as elastic lidars, except                                               
that Raman lidars feature a spectrometer-type receiver to discriminate the lidar 
returns according to wavelength.  Our receiver has 3 channels to detect the light 
intensity at various wavelengths. One channel collects the elastic lidar return, while 
the others correspond to the atmospheric oxygen and nitrogen Raman lidar backscatter, 
at about 375\,nm and 387\,nm, respectively.

The Raman lidar technique overcomes a major weakness of the elastic lidar measurements,  
in which the aerosol scattering contribution is present both in the 
transmission and backscattering components.  Extracting the aerosol attenuation from 
elastic lidar data therefore requires some assumptions regarding aerosol optical 
properties, mainly the relationship between aerosol backscatter and extinction.
For the Raman lidar, the aerosol scattering contribution is only present in the
transmission, so the only assumption to be made is the wavelength dependence 
of the aerosol transmission coefficient.  If the Raman lidar returns are measured 
for several wavelengths simultaneously, for example for molecular nitrogen and oxygen, 
even the wavelength dependence can in principle be retrieved without any assumptions.
In addition, if the elastic signal at the laser wavelength is measured simultaneously,
a comparison between the elastic Rayleigh and inelastic Raman return can be used to give
an independent estimate of the aerosol content as a function of height.

A practical disadvantage of the Raman lidar technique is the small Raman molecular 
cross section.  As a consequence, the laser source has to be operated at high power
and care has to be taken not to disturb the FD operation.  Currently, Raman lidar 
runs at the Pierre Auger Observatory are only performed right before and after the
regular FD operation.

\subsection{Raman Lidar Hardware and Data Acquisition}

The Raman lidar system consists of a Nd:YAG laser (ULTRA CFR) manufactured by
Big Sky Laser Technologies, Inc.  The emitted pulses at 355\,nm have a line width 
of $<3\,\mathrm{cm}^{-1}$, a pulse energy of about 10\,mJ at a repetition rate of 
20\,Hz, and a pulse duration of about 6\,ns.  The full angle divergence is less
than 1\,mrad. The emitted laser pulses are directed in the atmosphere (zenith 
direction) with a $2^{\prime\prime}$ steering mirror, which has a pointing sensitivity 
of about 0.1\,mrad. 

The Raman lidar has to meet a few general requirements.  It must                                        
be capable of sampling a range of altitudes from the ground to 7000\,m, be optimized 
to detect weak Raman backscattering from atmospheric nitrogen and oxygen, and efficiently 
suppress background light and cross-talk between the different channels.         

These conditions impose restrictions on the choice of the receiving optics, 
and on the design of the dichroic beam splitters and interference filters to 
be used in the wavelength discriminating detectors.  They also impose the use of notch 
filters providing an additional suppression of the strong elastically backscattered 
light in the Raman channels.  

In the setup of the Raman lidar system, the backscattered light is collected by a 
zenith pointing telescope (50\,cm diameter parabolic mirror, focal length 150\,cm) 
through a 10\,m long multi mode optical fiber coupled to the telescope with an 
aspherical lens.  The fiber transports the return light to the 3-channels beam 
separator: a combination of beam splitters, interference filters, notch filters, 
and neutral density filters.

After passing the final stage of the beam separator unit, the light is collected by 
Hamamatsu R1332 photomultipliers.  The voltage pulses at the output of 
the photomultipliers are amplified by a fast pre-amplifier (EG\&G Ortec, model 535).
A discriminator (CAEN N224) reduces the dark counts and forms 30\,ns wide NIM pulses. These 
pulses are detected by a PXI-5620 digitizer (National Instruments, maximum sampling rate 
64\,Msample/s) usually in a 0.200\,ms time window (30\,km range). The number of pulses 
in each of 1000 time windows (200\,ns wide, corresponding to a distance resolution of 30\,m) 
are counted, stored and displayed in a personal computer, after summing over a number of
laser shots.  The maximum counting rate is kept below 10\,MHz so that pulse pile-up effects 
are negligible.  The data production is presently quite small: 576 kByte for 40 minutes of 
data taking.

Fig.\,\ref{f:raman} shows the typical returns in the elastic and nitrogen/Raman channels. 
Note that the presence of a cloud layer at about 4.8\,km height appears as an attenuation 
in the Raman lidar return, since, as described above, the aerosol scattering is only present 
in the transmission.  In the elastic lidar signal, on the other hand, the cloud appears as an
increase in the signal due to the enhanced backscattering.   

The Raman lidar is remotely operated, sharing some services with the elastic lidar already 
operating at Los Leones.  Currently, the main goal of the Raman lidar is to measure the 
vertical aerosol optical depth.  Since the Raman measurements have to be taken outside 
FD operations, the measurement obtained with this technique is not directly usable for 
the analysis of FD data.  However, it can be used to cross-check and validate the
estimates of the aerosol content made by the subsequent elastic lidar operation.

\section{Conclusions and Outlook}\label{sec:conclusion}

Since March 2006, three of the four lidar stations are routinely operated each
night of FD data taking by a designated lidar shift taker at the central campus.
The fourth site is currently under construction and will start data taking in 
2007.  The three operating lidars perform continuous and discrete scans of the local 
atmosphere, and the data are analyzed for cloud coverage, cloud height, and the optical 
depth of cloud layers.  An analysis of the aerosol scattering and absorption parameters
will be added in the near future.

As the part of the lidar program that has the largest potential to interfere with the 
FD operation, the shoot-the-shower program is still in a testing phase.  Algorithms
to maximize the efficiency of StS are currently under development.  This includes
the implementation of software to filter background events that can erroneously
trigger the StS, such as lightning.

As a possible future extension of the elastic backscatter lidar program, an inelastic
Raman lidar is operated at the beginning of each data taking night.  

Another future upgrade currently under consideration is the installation of new 
mirrors with a larger focal length to reduce the speed of the lidar optics.  The large 
field of view of the current system implies that the background rate is high and the 
laser beam enters the field of view of the mirrors very early.  To lower the amount of
background light and overcome saturation at short range, a segmented mirror with a 1\,m
diameter and 1.1\,m focal length is currently under development.

\ack
We are grateful to the following agencies and organizations for financial support:
Comisi\'on Nacional de Energ\'{\i}a At\'omica and Gobierno de la Provincia de Mendoza, 
Argentina; INFN (Istituto Nazionale di Fisica Nucleare) (Italy); Slovenian Research 
Agency; National Science Foundation (USA) (contract numbers NSF-PHY-0500492, 
NSF-PHY-0134007); Department of Energy (DOE) Office of Science (USA) (DE-FG03-92ER40732).

\end{document}